\documentclass[twocolumn,aps,prb,showpacs,superscriptaddress,floatfix,citesort]{revtex4}
\usepackage{epsfig}  

\newcommand{\gl}[1]{Eq. (\ref{#1})}

\def\gtrless{\raise2.5pt\hbox{$>$}\llap{\lower2.5pt\hbox{$<$}}}
\def\gtrapprox{\raise2.5pt\hbox{$>$}\llap{\lower2.5pt\hbox{$\approx$}}}
\newcommand{\bsq}[1]{\begin{subequations}\label{#1}}
\newcommand{\esq}{\end{subequations}}
\newcommand{\beq}[1]{\begin{equation}\label{#1}}
\newcommand{\eeq}{\end{equation}}
\newcommand{\beqa}[1]{\begin{eqnarray}\label{#1}}
\newcommand{\eeqa}{\end{eqnarray}} \newcommand{\fur}{\qquad\mbox{for }\, }

\newcommand{\gd}{\dot{\gamma}}

\newcommand{\eps}{\varepsilon}
\renewcommand{\rho}{\varrho}

\newcommand{\add}[1]{#1}
\newcommand{\rem}[1]{}

\begin{document}

\title{Flow curves of dense colloidal dispersions: 
schematic model analysis of the shear-dependent viscosity near the
colloidal glass transition}

\author{Matthias Fuchs}
 \email[Corresponding author. Email: ]
  {matthias.fuchs@uni-konstanz.de}
\affiliation{Fachbereich Physik, Universit\"at Konstanz,
 78457 Konstanz, Germany
}

\author{ Matthias Ballauff}
\affiliation{Physikalische Chemie I, Universit\"at Bayreuth,
95440 Bayreuth, Germany}

\date{\today}


\begin{abstract}
\rem{The viscosity of a concentrated colloidal dispersion decreases
strongly with increasing shear rate ('shear thinning'). At the highest
volume fractions colloidal suspension may even behave as a yielding
solid.}  A recently proposed schematic model for the non--linear
rheology of dense colloidal dispersions is compared to
flow curves measured in suspensions that consist of thermosensitive
particles. The volume fraction of this purely repulsive model system
can be adjusted by changing temperature.  Hence, high volume fractions
($\phi \leq 0.63$) can be achieved in a reproducible manner. The
quantitative analysis of the flow curves suggests that the theoretical
approach captures the increase of the 
low shear viscosity with increasing density, the shear thinning for
increasing shear rate, and the yielding of a soft glassy solid. 
Variations of the high
shear viscosity can be traced back to hydrodynamic interactions which
 are not contained in the present approach but can be 
incorporated into the data analysis by an appropriate rescaling. \rem{The
comparison of theory and experiments demonstrates for the first time
that all features seen in the flow curves of concentrated suspensions
can be explained by the mode-coupling approach to the
colloidal glass transition. }
\end{abstract}
\maketitle

\section{Introduction}

The flow behavior of concentrated suspensions under steady shear is a
classical subject of colloid physics
\cite{russel,larson}. A large number of
experimental studies conducted mostly on hard spheres has 
established the basic facts: If the concentration of particles is not
too high, a \textit{first Newtonian region} is observed if the shear
rate $\dot{\gamma}$ is small. Here, the solution viscosity
  $\eta_0$ measured in this first Newtonian
 regime can be significantly larger than $\eta_{s}$ the one of the
  pure solvent. At higher shear rates, the perturbation of the
microstructure of the suspension by the convective
forces can no longer be restored by the Brownian motion of the
particles. Hence, significant \textit{shear thinning} will result in
which the reduced viscosity ${\eta}/{\eta_s}$ is more and more lowered
until (often) a \textit{second Newtonian region} is reached. In this
region, 
${\eta}/{\eta_s}$ is generally considered to be dominated
by the hydrodynamic interactions between the particles. Highly
concentrated suspensions behave as weak amorphous solids and
elastically withstand small but finite stresses \cite{Petekidis04}.

The experimental results obtained so far have demonstrated that the
deviation from the equilibrium structure can be gauged in terms of the
\textit{bare Peclet number} Pe$_{0} =
{a^2\dot{\gamma}}/{D_0}$, where $a$ denotes the particle radius
  and $D_{0}$ the diffusion coefficient at infinite dilution
 \cite{russel}. However,
non-Newtonian flow 
behavior is observed already at rather small Pe$_{0}$ and the
disturbance of the microstructure sets in at a shear rate defining a
second characteristic number, the Weissenberg number or dressed
Peclet number Pe=$\gd \tau$, which is connected to the structural
relaxation time $\tau$. 
Shear thinning may be considered to arise for Pe $\gtrsim 1$ . 
In dispersions able to order under shear, the viscosity is observed to
decrease. Yet, shear--thinning in concentrated 
suspensions of colloidal particles at low $\gd$
is not necessarily related to the onset of crystallization or other
effects solely occuring at very high shearing fields. The structure
remains amorphous during the application of  
shear rates that already lead to a marked decrease of the shear
viscosity \cite{Petekidis04,Laun92,Petekidis02b}.

The enormous raise of the zero-shear viscosity ${\eta_0}$
with increasing volume fraction has been a long-standing problem in
the field \cite{russel}. Earlier theoretical approaches 
\cite{Brady93,Lionberger00} have assigned this increase of
${\eta_0}/{\eta_s}$ to 
the onset of the structural arrest if the system is approaching the
volume fraction of random close packing located at ca. $\phi =
0.63$. Hence, ${\eta_0}/{\eta_s}$ is predicted to diverge at this
limit. However, Meeker et al. in 1997 \cite{Meeker97} carefully
re-analyzed all experimental data of ${\eta_0}/{\eta_s}$ available at
that time and concluded that the strong raise of the zero-shear
viscosity is related to the glass transition in suspension occurring
at the volume fraction $\phi_g \approx 0.58$. Other comparisons
strongly supported this view \cite{Goetze99}, while recent viscosity
measurements remained inconclusive \cite{Cheng02,FaradayDisc}. 

N\"agele and coworkers worked out a theoretical approach  
\cite{Banchio99,fuchsmayr} that could 
explain the increase of 
${\eta_0}/{\eta_s}$ on the base of the mode coupling theory (MCT) of
G\"otze and coworkers \cite{Goetze91b,gs}. 
Hence, the marked slowing down of the mobility
of concentrated suspensions can directly be traced back to the caging
of a given sphere by its surrounding neighbors. The quantitative
description of the dynamics of quiescent suspensions in terms of the
MCT has met with gratifying success when confronted with experimental
data obtained through dynamic light scattering
\cite{Megen93,Megen93b,Megen94,Beck99,Bartsch02,Eckert03}. 
Indeed, MCT was shown 
to explain the structural arrest of concentrated suspensions and
describe quantitatively the dynamics that stretch out over many orders
of magnitude. Moreover, in an important paper Mason and Weitz
\cite{Mason95} could demonstrate that MCT leads to a full explanation
of the linear viscoelastic behavior of hard sphere suspensions near the glass
transition. \\
Recently, a theoretical model for the shear-thinning of concentrated
suspensions was presented \cite{Fuchs02,Fuchs02b}. It is based on MCT
and gives a full description of the reduced viscosity
${\eta}/{\eta_s}$ as the function of the shear rate $\dot{\gamma}$. A
comparison with recent simulations has demonstrated that this theory
captures all the salient points of the flow behavior of glassy
systems \cite{Fuchs02c}. In particular, theory predicts a finite yield stress
beyond the glass point. It vanishes discontinuously when going
  below the glass transition, where a first Newtonian plateau appears,
  which is followed by strong shear thinning.  

  In this paper we present the
first comprehensive comparison with experimental data obtained from  a
model system. The paper is organized as follows: In section II we
review briefly the central concepts of our approach. 
 Section III then gives the quantitative comparison with recent
experimental data obtained on a model system \cite{Senff99}. A final
section will conclude this paper.

\section{Flow curves and the colloidal glass transition}

\subsection{Loss of structural memory caused by shear advection}

As mentioned above, the marked shear thinning in dense dispersions,
that is, the speeding up of structural relaxation through shear is not
necessarily related to shear ordering. This was shown by experiments
\cite{Laun92,Petekidis02b} as well as by Brownian 
dynamics simulations \cite{Strating99}.  In Refs. \cite{Fuchs02,Fuchs02b} we argued that
the speed up of decorrelation brought about
by shear advection combined with local Brownian motion lies at the origin
of shear thinning in dense dispersions. In this contribution we work out
the involved loss of structural memory caused by shearing in a schematic model
that captures the universal aspects of the full microscopic approach
of Ref. \cite{Fuchs02}. \\
The approach of Refs. \cite{Fuchs02,Fuchs02b},
 connecting  the nonlinear rheology of dense dispersions 
to the glass transition, predicts a transition from a
shear--thinning fluid to a yielding solid.
Even though small shear rates  are considered and the (bare) Peclet
number Pe$_{0}=\gd a^{2}/D_{0}$  is  negligible, the final relaxation of 
transient density fluctuations or of the transient stress moduli is
strongly accelerated by shear whenever Pe$=\gd\tau$ is not negligible.
In fluid states, where  $\tau$ is large at $\gd=0$, shear advection
speeds up the decay of structural correlations.  
 For states which would be solid without shear, and where the shear modulus
$G(t)$ would
 arrest at a (finite) elastic constant at  long
times,  enforcing  stationary shear leads to a finite relaxation time 
\cite{Fuchs02,Fuchs02b} which is of the order of $|\gd|^{-1}$. 
Hence, the glassy state of the suspension  is
shear--melted. The suspensions yields as stress fluctuations decay to
zero with rate set by the external drive. 

\subsection{Universal aspects}

The flow curves $\sigma$ versus $\gd$ exhibit qualitative aspects 
that are solely determined by the nature of the transition.
 With the separation parameter $\eps$
denoting the (relative) distance from the transition, and $t_0$ the
time scale obtained by matching onto  microscopic short--time motion,
the following behaviors of the steady state shear stress
 $\sigma$ in the `structural window' have been established \cite{Fuchs02c}
\beq{univ1}
\sigma= \sigma(\gd t_0,\eps) \to\left\{\begin{array}{ll}
\gd t_0 \; (-\eps)^{-\gamma}  G_{\infty}^{c} & \eps \ll - |\gd
t_{0}|^{1/\gamma}\\
\sigma^+_c \; \left( 1 + c_3 |\gd t_0 |^m \right) & |\eps| \ll |\gd
t_0|^{\frac{2a}{1+a}} \\
\sigma^+_c \; ( 1 + c_4 \sqrt{\eps} ) &  \eps \gg |\gd
t_0|^{\frac{2a}{1+a}} \end{array}\right. \; ,
\eeq
where the appearing constants are positive material--dependent
parameters and the 
exponents $\gamma$, $a$, and $m$ are non--universal numbers that are uniquely
determined by the quiescent static structure factor \cite{Franosch97,Fuchs02b}.
The first line of \gl{univ1}, which describes
the divergence of the viscosity, is familiar from classical MCT, and is
discussed in e.g. Ref. \cite{Fuchs04b}.
The `structural window', here, is defined as the double regime
$|\eps| \ll 1$ and $|\gd t_0| \ll 1$, where the slowing--down of the
structural dynamics dominates the steady state stress. 
A `dynamic yield stress' $\sigma^+(\eps)=\sigma(\gd\to0,\eps\ge0)$ is obtained
in the glass because a finite stress has to be overcome in order to
force the glass to yield even for vanishingly small shear rate.
The given asymptotes are only the
leading orders for $\eps\to0$ and $\gd t_0\to0$, while corrections
can be obtained systematically \cite{Franosch97,Fuchs02b}. A model calculation will be shown further below.

\subsection{Schematic models}

The universal phenomena summarized in \gl{univ1} exist in any model that exhibits
the bifurcation scenario from yielding solid to shear thinning fluid. The central feature of the  equations of motion is that
they contain the competition of two effects: i) a non-linear
memory effect  increases with increasing particle
interactions (`collisions' or `cage effect') which leads to a
non--ergodicity transition in the absence of shear, and ii), memory
effects vanish with time because of shear-induced decorrelation. Both 
effects can be captured in the simpler `schematic'
models also. Note that the models can be set-up so that they obey
similar stability equations as the microscopic approach. Thus, the
corresponding asymptotic results summarized in \gl{univ1} hold.\\

The well studied and comparatively simple schematic
F$_{12}^{(\gd)}$--model considers one normalized correlator $\Phi(t)$,
 which obeys a generalized relaxation equation \cite{Fuchs02b}:
\beq{sm1}
\dot{\Phi}(t) + \Gamma  \left\{
\Phi(t) +  \int_0^t\!\!\! dt'\; m(t-t') \, \dot{\Phi}(t')
\right\} = 0 \; .
\eeq
Without memory effects, $m\equiv0$, the correlator relaxes
exponentially, $\Phi(t)=\exp{-\Gamma t}$, but with $m\neq 0$,
retardation effects set in after a short--time variation (still given
by the initial decay rate $\Gamma$, viz.  $\Phi(t\to0)=1-\Gamma t+\ldots$). 
The correlator $\Phi(t)$ is taken to model the
normalized non--Newtonian shear modulus. 
A low order polynomial ansatz for $m$ suffices to model the feedback
 mechanism of the cage--effect.  We choose
\beq{sm2}
 m(t) = \frac{1}{1+(\gd t)^2} \; 
 \left(v_1 \Phi(t)+v_2 \Phi^2(t) \right) \; .
\eeq
Without shear, this model has been studied extensively
\cite{Goetze84,Goetze91b}. Increasing particle caging is modeled by
increasing coupling parameters $v_1$, $v_2 \ge0$, and the only effect of
shearing is to cause a time dependent decay of the friction kernel $m$.
The system loses memory because of shearing.
 The role of the transport coefficient
(viscosity) $\eta$ is played by the average relaxation time obtained
from integrating the correlator. It is also taken to determine
the stress: 
\beq{sm3}
\sigma = \gd\;  \eta = 
\gd \; \langle \tau \rangle = \gd \int_0^\infty\!\!\! dt\;  \Phi(t) \; .
\eeq
At high shear rates, the memory function is strongly supressed, so that
$\Phi$ returns to a single 
exponential, and the high shear visosity of the model
follows as $\eta_{\infty}=1/\Gamma$.

\subsection{Control parameter space and glass transition lines}

For the parameters of the model, the choice of generic values follows
from previous considerations \cite{Fuchs02b,Goetze91b}. 
First, the parameter $\Gamma$ sets the time scale and determines the short 
time dynamics. The bare Peclet number for the model thus is given by
Pe$_{0}=\gd/\Gamma$. This parameter hence can be compared directly to
the fast colloidal dynamics determined by the radius of the particles and
the short--time diffusion coefficient. 
Second, earlier studies suggest to choose the
two interaction parameters so that 
$v_2=v_2^c=2$ and $v_1=v_1^c+\eps/(\sqrt{v^{c}_{2}}-1)$, 
where $v_1^c=v_2^c(\sqrt{4/v_2^c}-1)\approx
0.828$. Thus, the decisive parameter,  namely the effective volume
fraction $\phi$ of the particles  enters the model only via
$\eps(\phi)$. A glass transition singularity lies at $\eps=0$, where the
long time limit $\Phi(t\to\infty)=f$ jumps from zero for 
$\eps<0$ to a finite value $f\ge f_{c}=1-1/\sqrt{v^{c}_{2}}$ for
$\eps\ge0$ and $\gd=0$. The parameter $f$ plays the role of the elastic
constant $G_{\infty}$ in this model. 

\begin{figure}[h]
\centering
\includegraphics[width=0.4\textwidth]{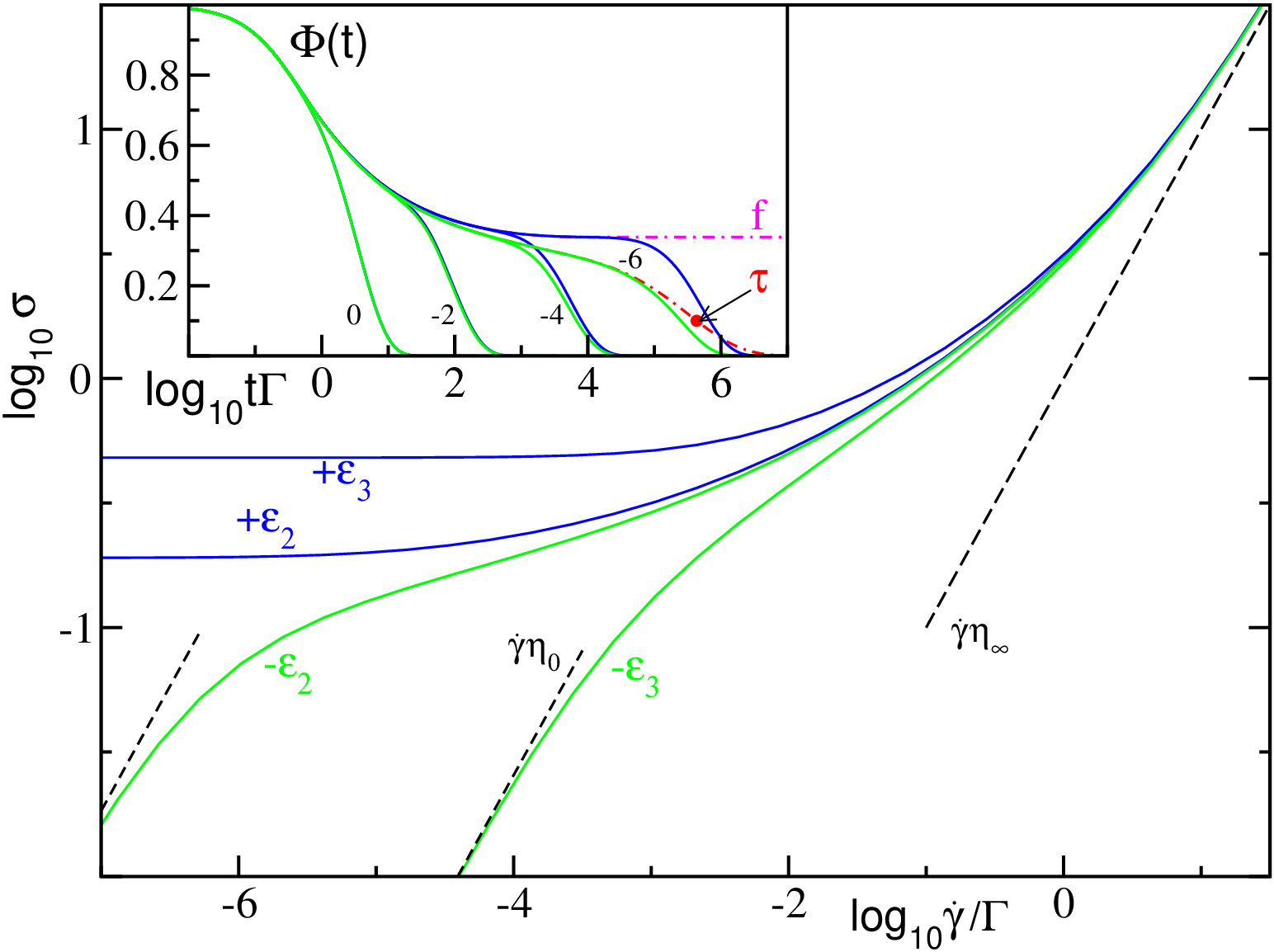}
\caption{Flow curves of the model, viz. dimensionless stress
$\sigma$ versus $\gd/\Gamma$, for two states that would be fluid respectively
 glassy without shear;  $\eps=\pm \eps_{2}$ and
 $\eps=\pm \eps_{3}$ (where $\eps_{3}=16 \,\eps_{2} = 0.0414$). 
 Straight lines with slope unity
 indicate the variation following the
low shear $\eta_{0}$ and high shear viscosity $\eta_{\infty}$. 
 \newline
The inset shows the correlators of the schematic
  $F_{12}^{{(\gd)}}$--model as function of rescaled time
 $t\Gamma$. The curve marked with a relaxation time $\tau$ taken
 at $\Phi(\tau)=0.1$ corresponds to a fluid state  without shear ($\eps=-\eps_{2}$, $\gd=0$). 
  The curve marked by the long time plateau value
 $f$ corresponds to a glass state without shear ($\eps=+\eps_{2}$, $\gd=0$).
  For increasing shear
 rates, $\log_{10}(\gd/\Gamma)=$ -6, -4, -2, 0 as labeled, the correlators decay
 more quickly at either value of $\eps$.  
\label{bild5h} }
\end{figure}

\subsection{Flow curves of sheared suspensions}

The presence of a glassy arrested structure is equivalent to 
a frozen in part in the correlator or memory function; 
thus without shear $\Phi(t\to\infty)=f>0$ and
$m(t\to\infty)=g>0$ hold for $\eps\ge0$. With
shear a non--decaying 
part in $m(t)$ is impossible, as $m(t\gd\gg1)\le (v_1+v_2)/(\gd
t)^2$; as a consequence, also $\Phi(t)$ always decays to zero. Memory is cut 
off at long times, and \gl{sm2} gives the most simple ansatz
recovering this effect 
of shear advection in the microscopic equations \cite{Fuchs02}, and
the obviously required symmetry in 
$\gd$.  The inset of Figure \ref{bild5h} shows the correlator for
fluid and glassy states for systems at rest ($\gd=0$), and for sheared
suspensions. For the latter systems, the $F_{12}^{(\gd)}$--model
predicts the speed up of the relaxation caused by increasing  
shear rates. Integrating over the correlators, as given in \gl{sm3},
leads to the viscosity  
which consequently exhibits shear thinning. The corresponding
 flow curves are shown in Fig. \ref{bild5h}: Here the
shear modulus $\sigma$ is plotted as the function of the reduced flow
rate $\gd /  \Gamma$. Theory predicts an evolution from an (almost) Newtonian
fluid at weak coupling to a markedly non--Newtonian fluid at stronger coupling
corresponding, in experiments, to
higher volume fraction. This is seen from the characteristic S-shaped
dependence of $\sigma$ on $\gd$. At the glass transition, there
is a discontinuous transition from the dissipative fluid--like
behavior to a yielding solid. 
A finite shear rate leads to a shear-melting of the glassy state.  

Fig. \ref{bild5h} presents the central result of theory. It gives the
full scenario for the non-linear flow behavior of dense suspensions
and relates it to the glass transition in these systems. Moreover, it
predicts that flow curves obtained from glassy suspensions should
present meaningful results. This is due to the fact that the shear is
expected to speed up the relaxation even at highest volume
fractions. As an experimental consequence of this, no hysteresis is
expected and flow curves present a well-defined
probe of the dynamics of glassy systems.  \add{As a caveat, though, the
condition needs to be recalled that the system is given enough time
to reach the steady state, and  that phase transitions and ordering
phenomena are prevented.}

\section{Comparison of model calculations and experimental data from
  model systems}  

\subsection{Thermosensitive latex particles}

A meaningful experimental study of the flow behavior of suspensions
requires a system of particles that exhibit a rather small
polydispersity and a high stability in the respective suspension
medium. Moreover, the particles should interact in a purely repulsive
fashion. This requirement is certainly given for the classical hard
sphere suspensions used for the study of colloidal glasses so
far. However, as an additional requirement, it should be possible to
prepare suspensions having volume fractions up to $0.63$ in order to
explore the region beyond the volume fraction of the glass
transition. 

Recently, we showed that aqueous suspensions of
thermosensitive latex particles meet these requirements
\cite{Senff99,Deike01}. The particles consist of a solid core of
poly(styrene) of ca. 100 $nm$ diameter onto which a shell of
crosslinked poly(N--isopropylacrylamide) (PNIPA) chains is affixed. The
particles are suspended in water and the PNIPA network in the shell is
swollen at low temperatures (ca. 10$^0C$). Raising the temperature
leads to an expulsion of this thermosensitive shell and the particles
will shrink. The advantage of the thermosensitive
suspension is obvious: The effective volume fraction could be changed
over a wide range by raising or lowering the temperature while keeping
constant the weight fraction of the particles. In this way highly
concentrated suspensions could be generated in situ, that is, directly
in the rheometer. No pre--shear was imposed on the system
by handling or filling in a highly concentrated suspension. Moreover,
any previous history of the sample could easily be erased by lowering
the effective volume fraction throuh raising the temperature. The
thermosensitive suspensions hence present a novel model system by
which volume fractions around and beyond the volume fraction of the
glass transition become accessible without freezing--in non--equilibrium
states caused by handling the suspension. Indeed, as shown in
Ref. \cite{Senff99}, the shear viscosity $\eta$ of 
suspensions of these particles could be obtained over a large range of
shear rates $\dot \gamma$ in a well-defined manner.  

It should be noted that the thermosensitive particles present a
well-studied system by now: The phase transition within the
microscopic network 
\cite{Seelenmeyer01} was shown to be fully reversible
\cite{Senff99,Deike01}. Moreover, an analysis of these particles by a
combination of small--angle neutron and X--ray scattering revealed that
the shell is well-defined and the particles exhibit a narrow size
distribution \cite{Seelenmeyer01,Dingenouts01}. Their interaction in
water is purely repulsive \cite{Deike01} if the temperature is not raised over
30$^0C$.

\subsection{Flow curves}

Theory states that flow curves of shear stress $\sigma$ versus shear
rate $\gd$ refer to a well-defined stationary state. Close to
vitrification, the  parameters characterising the static structure 
can be subsumised into the separation parameter $\eps$. Increasing the
interactions of the particles,  
brought about by increasing the volume fraction $\phi$ then is
described by increasing $\eps$ from negative values in the fluid to
zero, the point of the glass bifurcation,  and beyond, to positive
values in the glass. 

Equation (\ref{univ1}) demonstrates that there is in principle only one
parameter, the matching time $t_0$, that is required  
to determine the flow curves and other steady state averages in the structural
regime. This time scale contains all the effects of
hydrodynamic  interactions and other short--time phenomena not treated
by theory. Hence, $t_0$ needs to be adjusted by fitting the
theoretical 
curves to experimental  
data. Let us stress again, that in principle one time $t_0$ suffices to 
describe all different experimental measurements on a sheared
dispersion, for all volume fractions close to 
its glass transition, and for small shear rates. 

An 'idealized'
analysis of a complete experimental flow curve would thus proceed via
\beqa{analysis1}
\sigma &\to& \gd\; \eta_{\infty}
+ \sigma_{\rm struct.}\left(\eps , \gd t_{0}\right) \nonumber \\
&\approx&  \sigma_{\rm struct.}(\eps,\gd t_{0}) \fur
|\gd t_{0}|\ll1\,,\;  \; |\eps|\ll1\;,
\eeqa
where $\eta_{\infty}$ describes the flow curve at high shear
rates. At small shear rates, 
the two fit parameters $\eta_{\infty}$ and $t_{0}$ remain as unknowns for the
description of the structural region in the flow curve, as
$\sigma_{\rm struct.}(\eps,x)$ is determined by the static structure
at the transition (viz. the critical values $v_i^c$ of the vertices
in \gl{sm2}). Both parameters,
$\eta_{\infty}$ and $t_{0}$, are influenced by the
physics at high shear rates and short times, that is
dominated by hydrodynamic interactions.

\begin{figure}[h]
\centering
\includegraphics[width=0.4\textwidth]{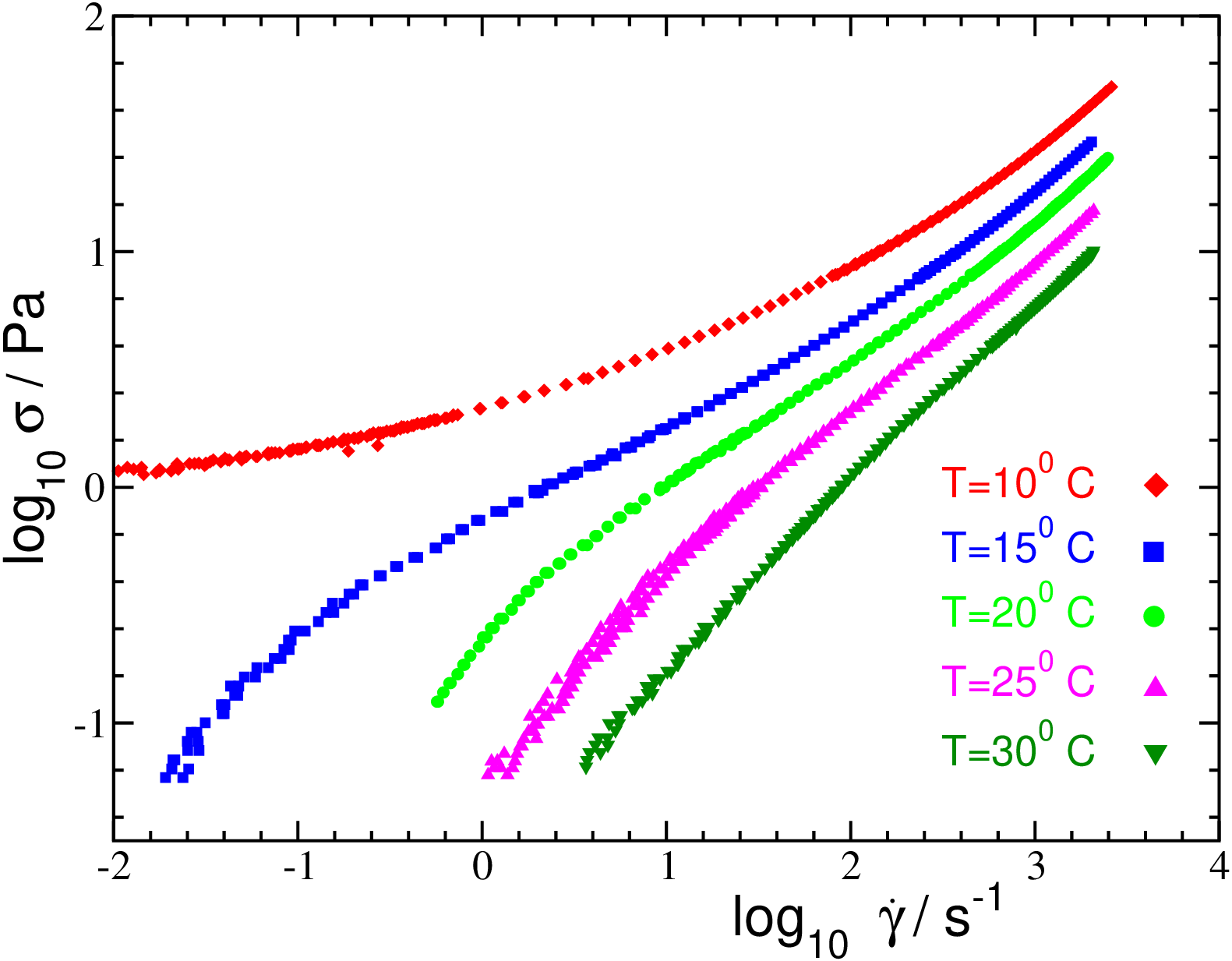}
\caption{Flow curves, viz. stationary
transverse stress $\sigma$ as function of shear
  rate $\gd$, for the thermosensitive latices close to glassy arrest
at various temperatures
  $T$ from top to bottom as denoted. The temperatures may be thought to
  correspond to different effective packing fractions. 
\label{bild6} }
\end{figure}

A representation where the subtleties of the non--linear flow curves
are revealed most clearly is given when plotting the stress versus
shear--rate. Figure \ref{bild6} shows the flow curves of the
thermosensitive particles for interaction strengths close to their
glass transition. The data have been taken from
Ref. \cite{Senff99}. Decreasing the temperature $T$ swells the
particles 
such that the effective packing fraction can be thought to
increase.  A region
for low $\gd$, where $\sigma$ depends strongly on density, can be
seen apart from one at high $\gd$, where a smaller variation is
found. It is the region at low shear rates which is treated by the
theory, i.e. where structural dynamics dominates the flow curves. Theory
suggests to plot the stress $\sigma$ 
versus shear rate $\gd$, instead of viscosity $\eta$ versus
$\gd$. This provides a direct comparison with Fig. \ref{bild5h}. The
  lack of 
straight pieces in Fig. \ref{bild6} indicates the absence of true
power--law shear--thinning, $\eta\propto\gd^{-x}$, which would show up
as $\sigma \propto \gd^{{1-x}}$.

\subsection{Elimination of effects from hydrodynamic interactions}

The ideal analysis of the structural part of
experimental flow curves, which only requires $t_0$ to be matched  at the
transition point, is hindered by $(i)$ inevitable
quantitative errors of the theory in calculating various constants, 
like e.g. the critical packing fraction, the yield stress at the
critical point, the transversal elastic constant, and other quantities that are
determined by the dispersion structure. 
Obviously, these quantities cannot be calculated within schematic
models where a small 
number of vertices $\{v_{i}\}$
replaces the structural information. In a schematic model analysis, therefore,
the overall stress amplitude has to be fitted; 
\beq{analysis2}
\sigma_{\rm struct.} =
\sigma_{0} * \sigma^{\rm theo.}(\{v_{i}\},\gd t_{0})\; .
\eeq
The time scale $t_{0}$ is easily determined within the schematic model,
and may then for convenience be eliminated in favor of the intrinsic
decay rate $\Gamma$ of the model. We adopt this convention, 
 as $\Gamma$ can be read off more easily from the $\sigma(\gd)$ curves. 

The ideal analysis must also require that  $(ii)$ 
 parameters like
 $\sigma_{0}$, $\eta_{\infty}$, and $t_0$ 
that are not treated by the model are constants. In particular,
 $\eta_{\infty}$, and $t_0$  are related to hydrodynamic
 interactions. To accomodate for these unknown 
parameters \add{and their non--negligible} density dependences, an
 analysis using schematic 
models can proceed via relaxing the restriction that all parameters
except for $\eps$ are constant. The data shown in Fig. \ref{bild6}
exhibit a  density dependence of the stress at high $\gd$ which is not
contained in the $F_{12}^{(\gd)}$--model; comparison with 
Fig. \ref{bild5h} shows that the model leads to a constant, 
$\sigma_{\infty}=\gd\eta_{\infty}=\sigma_{0} \gd/\Gamma$, which
thus does not constitute a full description of the hydrodynamic effects. 
Thus in \gl{analysis2}, a temperature dependence
needs to be included in the parameter  $\sigma_{\infty}(T)$. 
As discussed in Ref. \cite{Fuchs04b}, we expect that this needs to 
be accounted for by including a temperature dependence of $\sigma_0$.
In order to keep the number of temperature dependent parameters as small as
possible, in the following analysis all temperature dependent corrections 
beyond the model are
assumed to arise from $\sigma_{0}(T)$, the temperature dependence of
 the overall stress 
prefactor. The final expression used for the data analysis with the
$F_{12}^{(\gd)}$--model thus becomes
\beq{analysis3}
\sigma =
\sigma_{0}(T) * \sigma^{\rm theo.}(\eps(T),\gd/\Gamma)
\eeq
where, $\eps$ captures the temperature dependence of all vertex
parameters, as has been discussed repeatedly in the literature for
this model without shear.

\begin{figure}[h]
\centering
\includegraphics[width=0.4\textwidth]{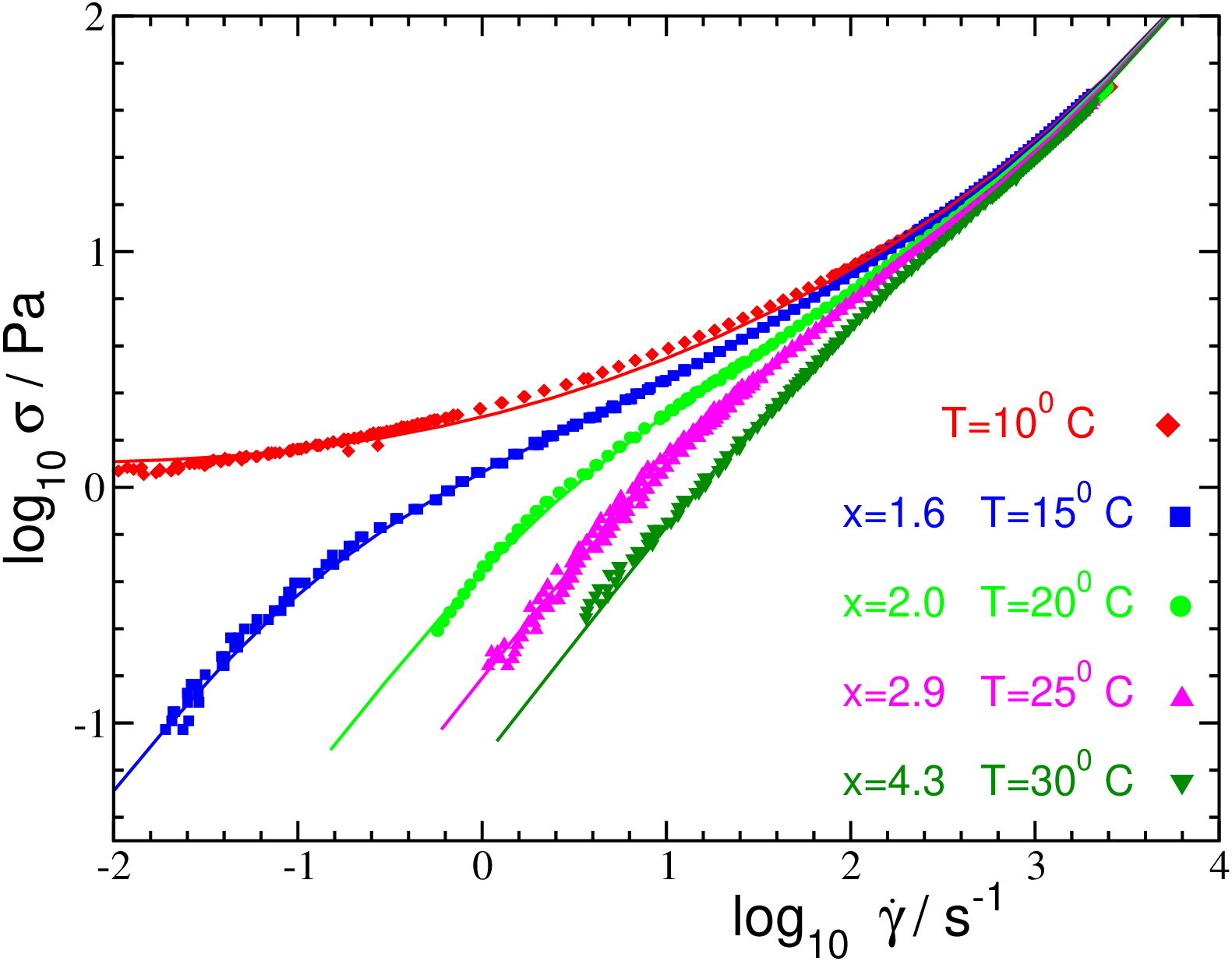}
\caption{Stress data from Fig. \ref{bild6}
 rescaled by  $x=\sigma_{0}(T=10^{0}C)/\sigma_{0}(T)$ to agree at
 high shear rate in order to eliminate corrections from i.a.
hydrodynamic interactions; rescaling  values $x$ from top to bottom
as indicated  in the legend. Solid lines give fits with the
  schematic model to the flow curves at small $\gd$ where the
  structural  dynamics dominates. The parameters are
 $\sigma_{0}(T=10^{0}C)=4.69$ Pa, $\Gamma=280 {\rm s}^{-1}$,
 for all curves, and separation parameters $\eps(T)$ as given in Table I.
The glass transition temperature lies close to $T_{c}=11^{0}C$.
\label{bild7}  } 
\end{figure}

\subsection{Data analysis}

The parameters obtained from fits of the schematic model curves
to the experimental data are given in table I.
The fits are shown in Fig. \ref{bild7}. Adjusting the parameters
starts with the data sets closest to the glass transition at
$\eps=0$, where the overall scales  $\sigma_{0}$ and $\Gamma$ can be
found as the flow curves show the strongest variations in curvatures,
and then proceeds to the temperatures farther away.
\begin{table}\label{tab:gl}
\begin{tabular}{|c||c|c|c|c|c|}
\hline
$T \mbox{ in} ^{0}C$ & 10 & 15 & 20 & 25 & 30\\
\hline
$\eps$ & 0.010 & -0.037 & -0.108 & -0.20 & -0.315 \\
\hline
$\sigma_{0}$ in Pa & 4.69 & 2.93 & 2.35 & 1.62 & 1.09\\
\hline
$\eta_{\infty}$ in $10^{-2}$Pa s & 1.67 &  1.04 & 0.84 & 0.58 & 0.39 \\
\hline
$\eta_{0}$ in  $10^{-2}$Pa s & $\infty$ & 341 & 27 & 5.7 & 1.7 \\
\hline
\end{tabular}
\caption{Parameters, $\eps(T)$ and $\sigma_{0}(T)$, for the fits of
  the flow curves in 
  Fig. \ref{bild7} using the $F^{(\gd)}_{12}$--model. The viscosities
  $\eta_{0}$ and $\eta_{\infty}$ are calculated from the fits using
  $\Gamma=280 s^{-1}$.} 
\end{table}

The transition to a yielding solid at $\eps(T_{c})=0$ is found to lie around
$T_{c}\approx 11^{0}C$, where the critical value of the yield stress
is $\sigma^{+}_{c}= 0.44$ Pa. It needs to be noted that the
  transition is not brought about directly by the change of
  temperature but by the change of the volume fraction of the
  particles with temperature. For lower temperatures, the yield
stress 
increases quickly, $\sigma^{+}(T=10^{0}C)=1.3$ Pa. At the fluid side of the
transition, the zero shear rate viscosity $\eta_{0}$ increases by more
than two decades, while the high shear rate viscosity $\eta_{\infty}$
changes by less than a factor 5. $\eta_{\infty}$ is not
predicted by the present model, it arises from hydrodynamic
interactions, but it is taken into account by by varying
$\sigma_{0}(T)$ with temperature.  

\begin{figure}[h]
\centering
\includegraphics[width=0.4\textwidth]{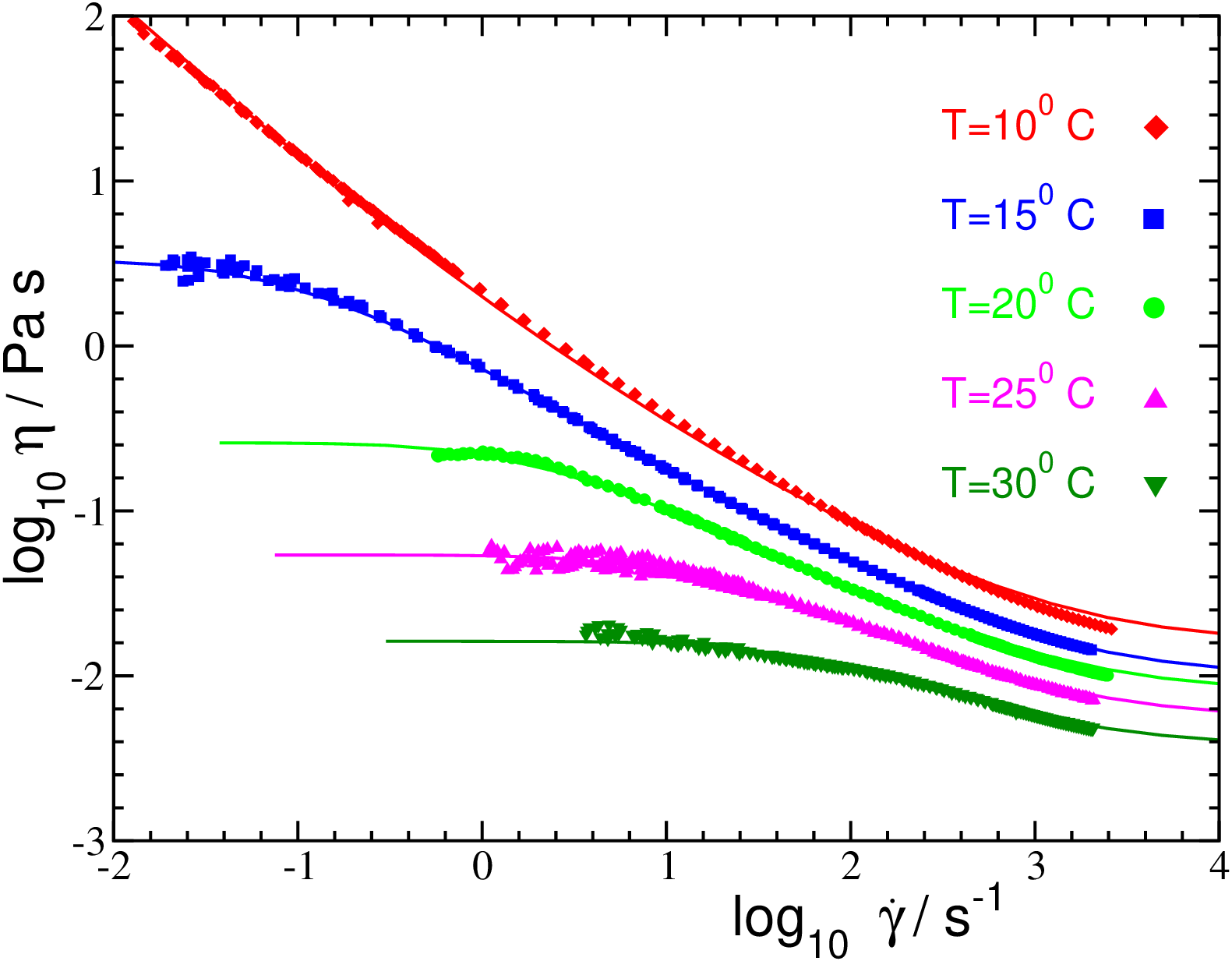}
\caption{ Viscosity data and corresponding schematic
  model fits taken from Fig. \protect\ref{bild7}. The second Newtonian
  plateau in the fit curves increases with lowering temperature
  because of the rescaling factors $x$.  Curves from top to bottom as labeled.
\label{bild8} }  
\end{figure}

Replotting the data and model results in the classical flow curves
showing  viscosity versus shear rate. This  
enlarges the variation along the ordinate and thus suppresses the
slight deviations of the fits from the data. Figure \ref{bild8} shows
the curves without rescaling so that the temperature dependence of the
high shear rate viscosity $\eta_{\infty}$ becomes apparent. 

\section{Conclusion}

A quantitative comparison of a first--principles approach to the
nonlinear rheology of dense colloidal systems
\cite{Fuchs02,Fuchs02b} with experimental data obtained in model
dispersions \cite{Senff99} has been given.  The employed schematic
$F^{(\gd)}_{12}$--model of the nonlinear rheology has been obtained
after simplifying steps based on the microscopic mode
coupling theory.  The speed up of the structural relaxation
brought about by shearing the suspension is the central mechanism
considered which causes shear thinning and yielding behaviors. It
enters the schematic model via a
time--dependent suppression of long--term memory. \\
The comparison with the experimental data demontrates that this model
  captures the essential features of the flow behavior of concentrated
  suspensions:  i) the strong shear thinning with increasing shear
  rate, and ii) at low shear rates the transition from a Newtonian
  liquid to a soft yielding solid.  Using the usual assumption that the hydrodynamic
  interactions can be described solely in terms of a high--shear
  viscosity $\eta_{\infty}$ a full description of the experimental
  data by Eq. (\ref{analysis3}) has become possible (see
  Fig. \ref{bild7}). The comparison demonstrates that the strong raise
  of the viscosity with increasing volume fraction can be fully
  explained by the structural arrest of the particles when approaching
  the volume fraction of the glass transition. It is hence evident
  that mode coupling theory that provides an excellent description of
  the dynamics of quiescent suspensions yields also a quantitative
  explanation of flow curves observed for
  suspensions subjected to a 
  steady shear field. 
 
\begin{acknowledgments}
We thank  
J.-L. Barrat, J. Bergenholtz, L. Berthier,
A. Latz and G. Petekidis for discussions. M.F. thanks M.E. Cates with whom the
        theoretical approach was developed for enlightening discussions.
 M.F.\ was supported by the DFG, grant Fu~309/3. 
M. B. was supported by the DFG, SFB 481, Bayreuth.
\end{acknowledgments}


\end{document}